\documentclass[twocolumn,nofootinbib,amsmath,amssymb,aps,prd,balancelastpage,superscriptaddress]{revtex4-1}

\usepackage{color}
\usepackage[active]{srcltx}
\usepackage{amsmath,amsfonts,amssymb,amsthm,amstext,amscd,eucal,srcltx}
\usepackage{epsfig,graphicx,bm}
\usepackage{epstopdf, epsf}
\usepackage{dcolumn}
\usepackage{hyperref}

\newcommand{\be}{\begin{equation}}
\newcommand{\ee}{\end{equation}}

\newcommand{\bse}{\begin{subequations}}
\newcommand{\ese}{\end{subequations}}
\newcommand{\bea}{\begin{eqnarray}}
\newcommand{\eea}{\end{eqnarray}}
\newcommand{\ba}{\begin{array}}
\newcommand{\ea}{\end{array}}
\newcommand{\bc}{\begin{center}}
\newcommand{\ec}{\end{center}}
\newcommand{\vect}{\mathbf}
\begin{document}
\preprint{IPM/P-2012/009}  
\vspace*{3mm}
\title{UV self-completion of a theory of Superfluid Dark Matter}%

\author{Andrea Addazi}
\email{andrea.addazi@lngs.infn.it}
\affiliation{Center for Field Theory and Particle Physics \& Department of Physics, Fudan University, 200433 Shanghai, China}

\author{Antonino Marcian\`o}
\email{marciano@fudan.edu.cn}
\affiliation{Center for Field Theory and Particle Physics \& Department of Physics, Fudan University, 200433 Shanghai, China}

\begin{abstract}
\noindent
We show that the model of superfluid dark matter developed in Refs.~\cite{Khoury:2014tka,Berezhiani:2015bqa,Berezhiani:2015pia}, which modifies the Newtonian potential and explains the galactic rotational curves, can be unitarized by the formation of classical configurations in the scattering amplitudes. The classicalization mechanism may also trigger the formation of the superfluid state from the early to the late Universe.  
\end{abstract}

\maketitle

\section{Introduction}
\noindent
A new paradigm for cold dark matter was recently suggested in Refs.~\cite{Khoury:2014tka,Berezhiani:2015bqa,Berezhiani:2015pia}. It was shown that the galactic velocity curves can be explained by the presence of a superfluid the phonons of which are weakly coupled to baryonic matter. A hybrid scenarios of cold dark matter and MOND (Modify Newtonian Potential) then arose. At the same time, it was shown that, contrary to the MONDs proposals, this kind of ``MOND superfluid dark matter'' is compatible with very constraining observations from clusters (in particular the Bullet cluster). 

The model is based on a non-relativistic effective Lagrangian, which {\it a posteriori} appears to be a phenomenologically healthy framework. But issues regarding the UV completion of the model have not yet solved in the literature. In particular, it was shown that the relativistic completion of the effective non-relativistic Lagrangian is a non-perturbative model with complicated highly non-linear interactions involving derivatives and polynomial mixed terms. 

Another unsolved aspect, which should be connected to the previous one, concerns the genesis of MOND superfluid dark matter. While in the traditional WIMPs or axion cold dark matter paradigms, the genesis mechanism is well known, in MOND superfluid is still unclear how such a state of matter could emerge in the early Universe. 

In this letter, we show that the model developed in Refs.~\cite{Khoury:2014tka,Berezhiani:2015bqa,Berezhiani:2015pia} exhibits a {\it non-Wilsonian} UV behavior. In particular, we show that the model belongs to the class of theories that are unitarized by the production of classical non-perturbative configurations in the scattering amplitudes. This is a mechanism dubbed {\it classicalization} by {\it Dvali and Gomez}, which was largely studied in many non-linear models \cite{Dvali:2010jz,Dvali:2012mx,Bajc:2011ey,Percacci:2012mx,Vikman:2012bx,Addazi:2015ppa}. In MOND superfluid model, the UV cutoff appears just around $1\, \rm meV$. This means that the theory predicts the production of classical configurations just around $1\, \rm meV$. 
This should also shed new light on the superfluid dark matter genesis mechanism. As shown in \cite{Khoury:2014tka,Berezhiani:2015bqa,Berezhiani:2015pia}, from best fits of galactic rotational curves, the phonon mass should be around ${1\div 10}\, \rm eV$ \footnote{It is tempting to suggest that the superfluid state is a neutrino superfluid, in which the condensation mechanism is triggered by a hidden non-linear electrodynamics sector coupled to neutrinos. This mechanism also connects the generation 
of the neutrino mass with the cosmological constant \cite{Alexander:2009yb,Addazi:2016oob}. }. Light Particles composing the superfluid can be generically produced as hot dark matter in the early Universe, arising from decays after the inflaton reheating. Their density ratio over baryonic density is directly controlled by their coupling with the inflaton --- or with other intermediate particles in which the inflaton decays.  With these assumptions, these particles should scatter at energies always much higher than the cut-off scale. Consequently, classicalons are formed in each scattering until the late Universe, when the thermal bath lower to $T\simeq 1\div 10^{-1}\, \rm meV$. These processes can be then interpreted as the formation of the superfluid state. For $E>\!\!>1~ {\rm eV}$, the formation and evaporation of classicalons may produce many soft particles, which partition the center of mass (CM) energy into a large number of emitted particles. Emitted particles should then be so soft to easily undergo a Bose-Einstein condensation mechanism. On the other hand, when the CM energy is above the threshold $E_{CM}\geq {\rm few}\, {\rm eV}$, the classicalon cannot evaporate into many particles, and states that are produced have a very tiny kinetic energy. Thus, the unitarization mechanism from classicalization may also be claimed to catalyze the production of the superfluid dark matter state. 

We emphasize that the appearance of light thermal relict particles can be made consistent with observations, only because of the mediation provided by classicalization effects. Indeed, as shown in Ref.~\cite{Berezhiani:2015bqa}, the light particles instantiating the superfluid dark matter picture shall be emitted through the mechanism of axion-like vacuum displacement. But the latter necessarily requires a temperature scale that is dependent of the parameter $\Lambda$ introduced in Ref.~\cite{Berezhiani:2015bqa}. We further notice that the details of classicalization are independent on the higher orders of the derivatives involved in the definition of the specific Lagrangian. Our analysis of the model introduced in Ref.~\cite{Berezhiani:2015bqa} shall not then be considered as the only working example of classicalizing theory, as we will specify in the next section.

 The paper is organized as follows. In Sec.~\ref{classic} we summarize the main ideas behind the concept of classicalization. In Sec.~\ref{mod} we introduce the superfluid cold dark matter model, as developed in Refs.~\cite{Khoury:2014tka,Berezhiani:2015bqa,Berezhiani:2015pia}. In Sec.~\ref{cas} we describe the production of the superfluid state from a cooling cascade, focusing on the energy partition processes. In Sec.~\ref{bo}, taking into account the thermal history of the Universe, we describe the phase transition to the superfluid state. Finally, in Sec.~\ref{co}, we spell our conclusions and discuss the novelty of this analysis. 

\section{Classicalization in a nutshell} \label{classic}
\noindent
 In order to understand the phenomenon of classicalization, we take into account as in Ref.~\cite{Dvali:2010jz} a real scalar field $\phi$ sourced by a current $J$ and governed by the theory 
\begin{eqnarray} \label{classex}
&&\mathcal{L}(\phi, J)=\mathcal{L}(\phi) + \frac{\phi}{M_*} J\,,\label{classex}
 \\ 
&&\mathcal{L}(\phi)=(\partial_\mu \phi)^2 + \frac{\phi}{M_*} \, (\partial_\mu \phi)^2 + \frac{\phi}{M^5_*} \, (\partial_\mu \phi)^4 + \dots \,. \nonumber
\end{eqnarray}
The non-linear terms in $\mathcal{L}(\phi)$ provide an example of the self-sourcing of the $\phi$ field. At the perturbative level, these terms entail violation of unitarity at CM energies $\sqrt{s}\!>\!\!>\! M_*$, but at the non-perturbative level the same admit restoration of unitarity via classicalization. 

For the theory in Eq.~(\ref{classex}), it is straightforward to show how classicalization takes place. As customary, we consider a source localized within a region of size $L$, with a strength of positive power in $1/L$.  This is the case, for instance, of energy-momentum type sources, as well as the case of the cubic self-coupling in Eq.~(\ref{classex}), and is enough to reproduce classicalization from scattering of $\phi$ particles at high energies. 

We may focus on the scattering of wave packets of $\phi$ at a CM energy $\sqrt{s}\!>\!\!>\! M_*$, and let the particles scatter with transferred momentum $1/L$, at a minimum distance $L$. As far as $L>L_*=1/M_*$, perturbatively wave-packets are dealt with as quantum states. However, the main idea underlying the analysis in  Eq.~(\ref{classex}) is that when wave-packets are localized within a distance $r_*=L_*^2\sqrt{s}$, the physical system enters a classical regime through the development of classical configurations, i.e. it classicalizes. In other words, self-sourcing implies that the localization of quanta of CM energy  $\sqrt{s}\!>\!\!>\!M_*$ cannot happen at distances shorter than $r_*=L_*^2\sqrt{s}$.

To prove classicalization one may consider a simple argument \cite{Dvali:2010jz}. Let us localize particles of CM energy $\sqrt{s}=1/L\!>\!\!>\!M_*$ within a sphere or radius $R\!<\!\!<\!r_*$. Cubic self-interactions trigger production of an effective localized source for $\phi$ that behaves like $J\sim M_*/L$. Consequently, linearizing the equation of motion for $\phi$ in the weak field regime entails the equation of motion $\Box \phi + \dots =\delta(r) L_*/L\,.$ The presence of an energy source for $\phi$ implies that a gaussian flux for $\nabla \phi$ is generated, and thus the radial dependence for $\phi$, i.e. 
$$\phi(r)\sim \frac {L_*}{L \,r}\,,$$
can be recovered. Consequently, approaching the localization region $\phi(r)$ increases, until the field becomes of order $M_*$ at a distance $r_*=L_*^2/L$. In other words, a source of energy $\sqrt{s}=1/L\!>\!\!>\!M_*$ extends over an effective distance $r_*\!>\!\!>\! L_*$, thus becoming a classical. This has profound consequences, which require to step out from the intuitions developed in the perturbative approach. In the latter framework, one might think that high-energy scattering happening at $\sqrt{s}\!>\!\!>\!M_*$ will be governed by hard-collisions, and that the momentum transfer will be proportional to $\sqrt{s}$. Instead, as agued above, the high-energy processes entail formation of classical systems of radius $r_*=L_*^2 \sqrt{s}$. These latter may then undergo a cascade of decays into many particles states. In other words, the $2\rightarrow 2$ particles scattering processes will be characterized by amplitudes that are very suppressed at very high transferred momenta, and dominated by soft transferred momenta interactions at $\sim 1/r_*$.

Classicalization, which is triggered by sources $\sqrt{s}=1/L\!>\!\!>\!M_*$, then happens at very high energies, but is characterized by a long-distance dynamics that corresponds to momenta $1/r_*\!<\!\!<\!M_*$. This implies that higher-order corrections are under control, and non-quadratic terms like the ones appearing in Eq.~\eqref{classex} do not change the dynamics of classicalization. To better show this, we consider the generic Lagrangian terms 
\begin{equation} \label{ho}
\frac{1}{M_*^{n+4k-4}} \phi^n (\partial_\mu \phi \partial^\mu \phi)^k\,, 
\end{equation} 
which generalize the ones appearing in Eq.~\eqref{classex}. Considering the asymptotic behavior at large $r$ to be $\phi(r)\sim1/(M_*L r)$, terms like in Eq.~\eqref{ho} contribute as
\begin{equation} \label{hoe}
\frac{1}{M_*^{n+4k-4}} \phi^n (\partial_\mu \phi \partial^\mu \phi)^k\,\sim\, \frac{1}{M_*^{2n+6k-4} L^{2k+n} r^{4k+n}} \,.
\end{equation} 
This allows to define the radius 
$$r_{n,k}=L_*\left( \frac{L_*}{L}\right)^{\frac{2k+n-2}{4k+n-4}}\,,$$ 
at which non-linear terms become of the same order of the linear term --- characterized by $n=0$ and $k=1$. 

One may immediately observe that $r_*$ represent the largest of the $r_{n,k}$ scales. In particular, assuming $k=1$ --- i.e. no higher-order kinetic terms appear in the Lagrangian --- the scales $r_{n,1}$ are all equal to $r_*$. The addition of higher order derivatives for $k>1$ can only lower the value of $r_{n,k}$, which would eventually become smaller than $r_*$. Nonetheless, lower $k$ terms will have to be present anyway in the Lagrangian considered, setting up to $r_*$ the classicalization scale. In other words, the scale $r_*$ is determined by the leading non-linear terms, the higher non-linear ones only affecting the dynamics of $\phi$ at distances $r\!<\!\!<\!r_*$. This argument is enough to ensure that the size of the $r_{n,k}$ radii cannot change the size of the classical configuration, and consequently cannot play an important role in scattering processes at large $s$.

\section{The Model} \label{mod}
\noindent
 In this section we show how the classicalization phenomenon can work for superfluid dark matter, using the same analysis' techniques introduced in Refs.~\cite{Dvali:2010jz,Dvali:2012mx}, and then deployed in Ref.~\cite{Bajc:2011ey}.

We start from the relativistic Lagrangian proposed in Refs. \cite{Berezhiani:2015bqa,Berezhiani:2015pia}, namely 
\begin{eqnarray} \label{L}
\mathcal{L}=&&-\frac{1}{2}\left(|\partial_{\mu}\Phi|^{2}+m^{2}|\Phi|^{2}\right) \nonumber\\
&&-\frac{\Lambda^{4}}{6(\Lambda_{c}^{2}+|\Phi|^{2})^{6}}\left(|\partial_{\mu}\Phi|^{2}+m^{2}|\Phi|^{2} \right)^{3} \,.
\end{eqnarray}
The equation of motion in presence of a constant localized source term $Q\,\delta^{4}(x-y) \,\Phi$ entering the Lagrangian is 
\begin{eqnarray} \label{L}
Q\delta^{4}(x-y)=&&\left(\Box-m^{2}\right)\Phi \\
&&\!\!\!\!\!\!\!+12\partial^{\mu} \{U(\Phi)\left(|\partial \Phi|^{2}+m^{2}|\Phi|^{2} \right)^{2}|\partial \Phi|\} \nonumber\\
&&\!\!\!\!\!\!\!-2U'(\Phi)\left(|\partial\Phi|^{2}+m^{2}|\Phi|^{2}\right)^{3} \nonumber\\
&&\!\!\!\!\!\!\!-12m^{2}U(\Phi)\left( |\partial \Phi|^{2}+m^{2}|\Phi|^{2}\right)^{2}\Phi \nonumber\,,
\end{eqnarray}
where 
\begin{eqnarray}
U(\Phi)\equiv \frac{\Lambda^{4}}{6(\Lambda^{2}+|\Phi|^{2})^{6}}
\end{eqnarray}
and $J$ is the source's current. 

The last term in equation (\ref{L}) has a perturbative expansion
\begin{eqnarray}
\frac{\Lambda^{4}}{\Lambda_{c}^{12}}\sum_{k=0}^{6}\frac{(-1)^{k}(k+5)!}{k! 5!}\left(\frac{|\Phi|}{\Lambda_{c}}\right)^{2k}
\left(|\partial \Phi|^{2}+m^{2}|\Phi|^{2} \right)^{3}\,. \nonumber
\end{eqnarray}
These are operators of the form 
\begin{eqnarray} \label{ope}
\frac{1}{M_{*}^{n+4k-4}}|\Phi|^{n}|\partial_{\mu}\Phi|^{2k} \,,
\end{eqnarray}
in which $\Lambda_{c}\simeq \Lambda$ and $M_{*}\simeq \Lambda \simeq 1\, \rm meV$ is assumed. In what follows we describe the scattering of quanta of the effective dark matter field, in first approximation, by the contact-interaction terms associated to the operators in \eqref{ope}. We then describe the asymptotic limit of the operators in terms of the energy of the quanta in the center of mass (CM) of the colliding system. The asymptotic behavior of $\Phi$ at large $r$ is $(M_{*}L)^{-1}r^{-1}$, where $L\simeq E_{CM}^{-1}$ and $E_{CM}$ is the CM energy. Operators then contribute asymptotically to scattering amplitudes as 
\begin{eqnarray}
(M_{*}^{2n+6k-4}L^{2k+n}r^{4k+n})^{-1}\,,
\end{eqnarray}
in which $L=E_{CM}^{-1}<\!\!<L_{*}$ and we have defined $L_{*}=M_{*}^{-1}$.

Notice that non-linear terms diverge at short-distances, in the limit $r\rightarrow 0$, and that in this limit derivative terms contribute as $L_{*}/r$. In the limit $r>\!\!>L_{*}$, these terms remain subdominant. When $M_{*}\simeq \Lambda$, derivative terms remain subdominant for distances of the order of the Hubble radius $(\Lambda \sim 1 \, \rm meV)$. 

We can now recast the Lagrangian \eqref{L} in terms of the fluctuations around the classical solution, namely 
\begin{eqnarray}
\Phi=\Phi_{cl}+\delta \Phi\,,
\end{eqnarray}
and then find the perturbed action
\begin{eqnarray} \label{Act}
\!\!\!\!\!\int \!\! d^{4}x[\mathcal{L}_{cl}-\delta \Phi Q \delta^{4}(x-y)\!+\!\frac{1}{2}\delta \Phi O \delta \Phi+\mathcal{L}_{int}(\delta \Phi)]\,,
\end{eqnarray}
where $O$ is the quadratic operator associated to the Lagrangian (\ref{L}), and $Q$ has been rewritten as a constant source $J$ localized in $y$. The operator $O$ has a complicated expression, but fortunately for our purposes, we may only consider the asymptotic forms of these operators in the large distance limit and in the short distance one. In the large distance limit, it is easy to show that the operator $O$ converges to 
\begin{eqnarray} \label{O1}
O(r\rightarrow \infty)\rightarrow -\left(\frac{\partial^{2}}{\partial r^{2}}+\frac{2}{r}\frac{\partial}{\partial r}+\frac{L^{2}}{r^{2}}\right)\,,
\end{eqnarray}
which corresponds to free-propagating scalars. In \eqref{O1} $L$ stands for the differential representation of the angular momentum operator. 

In the opposite regime, one obtains a series of polynomially divergent terms 
\begin{eqnarray} \label{O2}
O(r\rightarrow 0)\rightarrow \frac{1}{r^{4k+n}} \left(c_{1}\frac{\partial^{2}}{\partial r^{2}}+c_{2}\frac{2}{r}\frac{\partial}{\partial r}+c_{3}\frac{L^{2}}{r^{2}}\right)\,,
\end{eqnarray}
$c_{1,2,3}$ being numerical prefactors that are irrelevant for the following arguments, which are based on asymptotic limits. 

%


This suggests that the two asymptotic limits are interpolated by a classical solitonic solution, dubbed classicalon. In our case, to find an analytic solution is very challenging. Solutions can indeed be founded only numerically. However, in the case of a similar but simpler higher-derivative theory, with similar asymptotic behaviors, as a $\frac{1}{4}M_{*}^{-2}(\partial \Phi)^{4}$ theory, a classical solution can be found by imposing the following ansatz inside the equation of motion: 
\begin{equation}
\label{soll}
\partial_{\mu}\Phi_{cl}=\frac{(x-y)_{\mu}}{|x-y|}M_{*}^{2}\mathcal{F}\Big(\frac{|x-y|}{r_{0}(Q)} \Big)\,,
\end{equation}
with 
\begin{equation}
\label{radius}
r_{0}(Q)=\sqrt{\frac{Q}{M_{*}^{2}\Omega_{2}}}\, ,
\end{equation}
with $\mathcal{F}$ still generic. Plugging Eq.(\ref{soll}) into the EoM, one obtains for $\mathcal{F}$ the form
\begin{equation}
\label{plugging}
\mathcal{F}(\rho)=\Big(\sqrt{\frac{1}{27} + \frac{1}{2\rho^{6}}}  +\frac{1}{2\rho^{3}}   \Big)^{1/3}-\Big(\sqrt{\frac{1}{27} + \frac{1}{2\rho^{6}}}  -\frac{1}{2\rho^{3}}   \Big)^{1/3}\,,
\end{equation}
where $\rho=\frac{|x-y|}{r_{0}(Q)}$. 


The classical solitonic field associated to the classicalon can be expanded in plane waves as follows:

\begin{equation}
\label{plan}
\Phi_{cl}= \int \frac{d^{3}k}{\sqrt{(2\pi)^{3}2\omega_{k}}}\Big(e^{ikx}\alpha_{k}+e^{-ikx}\alpha_{k}^{*}\Big)\,.
\end{equation}
In the framework of such an expansion, the classicalon can be represented by a coherent state $|class\rangle$. The Fourier expansion coefficients of $\phi_{cl}$ are defined as the classicalon expectation values of the Fock space annihilation/creation operators 
 $\hat{C}_{k},\hat{C}_{k}^{\dagger}$, which read
\begin{equation}
\label{asolosol}
\langle class | \hat{C}_{k}|class\rangle =\alpha_{k},\quad {\rm for} \quad [\hat{C}_{k},\hat{C}^{\dagger}_{k'}]=\delta_{k,k'}\, . 
\end{equation}
Let us note that $\hat{C}$ and $\hat{C}^{\dagger}$ are not the annihilation/creation operators of the asymptotic propagating fields in the initial Lagrangian. In stead, they create and destroy classicalon quanta, which evidently have different dispersion relations then free fields. In general, a state $|class\rangle$ is composed of a tensor product of coherent states with different four-momenta $k$, namely
\begin{equation}
\label{jakal}
|\alpha_{k}\rangle=e^{-\frac{1}{2}|\alpha_{k}|^{2}}e^{\alpha_{k}\hat{C}^{\dagger}}|0\rangle=e^{-\frac{1}{2}|\alpha_{k}|^{2}}\sum_{n_{k}=0}\frac{\alpha_{k}^{n_{k}}}{\sqrt{n_{k}!}}|n_{k}\rangle\, , 
\end{equation}
where $|n_{k}\rangle$ is the Fock state with $n$-field of $k$-momenta. In this framework, the number operators is defined as $N=\int_{k}\alpha_{k}^{*}\alpha_{k}$.

The production of a classicalon configuration cannot be simply calculated from a $in\rightarrow out$ tree level S-matrix, where the $|in\rangle$ and $|out\rangle$ states are just asymptotic free field states. In fact, the production and evaporation of a classicalon configuration will generate a large number of entangled quanta, which must be in the same coherent state. In other words, the transition that must be evaluated for our purposes --- showing the unitarity of scatterings from classicalon production --- is the $in\rightarrow classical$ transition. Fortunately, such a transition can be easily related to an $in\rightarrow out$ S-matrix, calculable within the standard quantum field theories techniques, by the relation
\begin{equation}
\label{jaaj}
\langle in|S|class\rangle = \langle in|S|out\rangle \langle out|class\rangle\, . 
\end{equation}
In other words, we must just evaluate the transition element $\langle out|class\rangle$. 

Replacing $|out\rangle$ with any arbitrary number of particles with the same momenta, this transition element can be written as $|n(k)\rangle=(a^{\dagger}(k))^{n(k)}/(\sqrt{n(k)!}|0\rangle$ (where $a^{\dagger}$ is the free quanta creation operator). This implies 
that, for a generic $n(k)$, one obtains 
\begin{equation}
\label{jaa}
\langle n(k)| coherent\rangle=e^{-\frac{N}{2}}+{\rm subleading\,\,\, terms}\, . 
\end{equation}

We emphasize that from the perspective of the classicalon quanta, any $|out \rangle$ state is empty of classicalon quanta (created by $C^{\dagger}$). 

In the next subsection, we will show that the classicalon production lead to the unitarization of a sub-set of operators in the initial Lagrangian. Similar considerations can be extended to other operators with more technical difficulty issues.

\subsection{Scattering amplitude analysis}

\noindent 
To derive amplitudes in a more controllable way, we consider a $U(N_{f})$ global flavor symmetry. We introduce a multiplet of scalar real fields transforming in the adjoint representation $U(N_{f})$, namely $\Phi \equiv \Phi^{a}t_{a}$, and label with $t_{a}$ the symmetry generators of $U(N_{f})$, fulfilling $[t^{a},t^{b}]=i\sqrt{2}f^{abc}t^{c}$, with $f^{abc}$ the totally antisymmetric structure constants of $U(N_{f})$. We then consider the subset of operators, with flavor symmetry, 
\begin{equation}
\label{phin}
c_{n}(\partial \Phi)^{T}\cdot \frac{1}{M_{*}^{2n}}(s\{\Phi\})^{2n}\cdot (\partial \Phi)\,,
\end{equation}
where $c_{n}=(-1)^{n}(n+5)!/(n! 5!)$, corresponding to fixing $k=1$ in Eq.(\ref{ope}). The $s\{\Phi\}$ has an group-indices structure of the form 
\begin{equation}
s\{\Phi\}^{ab}=-if^{cab}\Phi^{c}\,, 
\end{equation}
while ``dot" in Eq.\ref{phin} denotes the contraction of indices in the adjoint representation.

It is worth emphasizing that this assumption does not change the low-energy analysis carried out by Berezhiani \& Khoury. 
In this flavored version of the model, every possible Feynman diagram can be automatically included through the powerful spinor helicity formalism. All the interaction vertices are controlled by the structure constants $f^{abc}$. Very much in the same way of Yang-Mills theories or pion amplitude models, it is then straightforward to demonstrate that any tree level on-shell amplitude has the simple group structure
$$\mathcal{M}^{a_{1}a_{2},..a_{n}}(p_{1},p_{2},...,p_{n})=$$
\begin{equation}
 \label{MauMd}
\sum_{\sigma \in S_{n}/Z_{n}}\langle t^{a_{\sigma(1)}} t^{a_{\sigma(2)}} ...t^{a_{\sigma(n)}} \rangle \mathcal{M}_{\sigma}(p_{1},p_{2},...,p_{n})\, ,
\end{equation}
where brackets $\langle \  \dots \  \rangle$ denote the Killing form on the group, i.e. the trace on the algebra elements. All the momenta are conventionally introduced with the incoming signature, and the sum is taken over $1,2,...,n$ permutations, modulo $Z_{n}$-cyclic permutations. The same structure of Eq.(\ref{MauMd}) can be obtained for generic interactions vertices $V^{a_{1},...a_{n}}_{n}(p_{1},...,p_{n})$. 
After some tedious but straightforward algebra, flavor ordered interaction vertices are found to have the form:
$$V_{2n+1}(p_{1},....,p_{n+1})=0\, , $$
$$V_{2n}(p_{1},...,p_{n})=$$
\begin{equation}
\label{Generic}
\frac{c_{n}}{M_{*}^{2n-1}}\sum_{k=1}(-1)^{k-1}\frac{(2n-2)!}{(k-1)!(2n+k-1)!}\sum_{i=1}^{2n}(p_{i}\cdot p_{i+k})\, . 
\end{equation}

Moving from the vertex structures, we can now calculate the flavor ordered amplitudes. The kinematic structure of the eight-point amplitudes, written in terms of the Mandelstam variables $s_{ij}=(p_i+p_j)^2$, then reads  

$$M_{*}^{6}\mathcal{M}(1,....,8)= a_{1}\frac{(s_{12}+s_{23})(s_{14}+s_{47})(s_{56}+s_{67})}{s_{13}s_{57}}+$$
\begin{equation}
+a_{2}\frac{(s_{12}+s_{23})(s_{14}+s_{45})(s_{67}+s_{78})}{s_{13}s_{68}}
\end{equation}
$$+a_{3}\frac{(s_{12}+s_{23})(s_{45}+s_{47}+s_{56}+s_{58}+s_{67}+s_{78})}{s_{13}}$$
$$+a_{4}s_{12}+a_{5}s_{14}+{\rm cycl}\, ,$$
where $a_{i}$ are numerical prefactors that are not relevant for our purposes. Similarly, the ten-point amplitude casts
$$M_{*}^{8}\mathcal{M}(1,..,10)=-\frac{s_{12}+s_{23}}{s_{12}}\Big\{$$
$$+b_{1}\frac{(s_{14}+s_{49})(s_{58}+s_{69})(s_{67}+s_{78})}{s_{59}s_{68}}$$
$$+b_{2}\frac{(s_{14}+s_{45})(s_{18}+s_{69})(s_{67}+s_{78})}{s_{15}s_{68}}$$
$$+b_{3}\frac{(s_{18}+s_{49})(s_{45}+s_{58})(s_{67}+s_{78})}{s_{48}s_{68}}$$
$$+b_{4}\frac{(s_{14}+s_{45})(s_{16}+s_{67})(s_{18}+s_{89})}{s_{15}s_{17}}$$
$$+b_{5}\frac{(s_{14}+s_{45})(s_{16}+s_{69})(s_{78}+s_{89})}{s_{15}s_{79}} $$
$$+b_{6}\frac{(s_{18}+s_{49})(s_{47}+s_{58})(s_{56}+s_{67})}{s_{48}s_{57}}$$
$$+b_{8}\frac{(s_{16}+s_{49})(s_{45}+s_{56})(s_{78}+s_{89})}{s_{46}s_{79}}   $$
$$+ b_{9}\frac{(s_{14}+s_{18}+s_{45}+s_{49}+s_{58}+s_{69})(s_{67}+s_{78})}{s_{68}}$$
$$-b_{10}\Big(\frac{(s_{18}+s_{49})(s_{45}+s_{47}+s_{56}+s_{58}+s_{67}+s_{78})}{s_{48}}$$
$$+\frac{(s_{14}+s_{16}+s_{45}+s_{47}+s_{56}+s_{67})(s_{18}+s_{89})}{s_{17}}$$
$$+\frac{(s_{14}+s_{16}+s_{45}+s_{49}+s_{56}+s_{69})(s_{78}+s_{89})}{s_{79}}$$
$$+\frac{(s_{14}+s_{45})(s_{16}+s_{18}+s_{67}+s_{69}+s_{78}+s_{89})}{s_{15}}$$
$$+\frac{(s_{14}+s_{49})(s_{56}+s_{58}+s_{67}+s_{69}+s_{78}+s_{89})}{s_{59}}\Big)$$
$$+b_{11}s_{14}+k_{12}s_{16}+k_{13}s_{18}+k_{14}s_{45}$$
$$+b_{15}s_{47}+k_{16}s_{49}+k_{17}s_{56}+k_{18}s_{58}$$
\begin{equation}
\label{jaja}
b_{19}s_{67}+b_{20}s_{69}+b_{21}s_{78}+b_{22}s_{89}\Big\}\,,
\end{equation}
where $b_{i}$ are numerical prefactors again not relevant for our purposes.

It is worth to note that both 8 and 10 point amplitudes are maximized if all final states have the same final energy, as a democratic redistribution of the CM energy on all of them.  Following the kinematical structure of these amplitudes, up to a generic $N+2$ large number of insertions, one can easily obtain the leading order amplitude
\begin{equation}
\label{amplitude}
M_{*}^{N}\mathcal{M}(1,2,...N+2)=K_{N+2}\, s+O(1/N)\,,
\end{equation}
where $K_{N+2}\sim O(1)$ is a constant not relevant for our purposes, and the CM energy $\sqrt{s}$ is democratically distributed into final state momenta. This implies that, after integrating on the phase-space, the corresponding cross section $2\rightarrow N$ will behave as follows
\begin{equation}
\label{MNNd}
\sigma=\sigma_{0} N!\Big(e^{2}\frac{s}{N^{2}M_{*}^{2}}\Big)^{N}\sim \sigma_{0}\frac{1}{(N-1)!}\Big(c  \frac{s}{M_{*}^{2}} \Big)^{N}\, ,
\end{equation}
where $\sigma_{0}$ is a $O(1)\, M_{*}^{-2}$ quantity. \\

To account for a final coherent state describing classicalons, we shall evaluate the CM energy $\sqrt{s}$ on the latter state, and then insert into Eq.~(\ref{MNNd}) the scaling factor determined in Eq.~\eqref{jaaj}. For relativistic particles, the energy scales as the momentum, and the expectation value on a coherent state of the momentum scales as $\sqrt{N}$. Consequently, $s$ will scale with $N$ --- this property of classicalons was also discussed in details in Ref.~\cite{Dvali:2012en} --- and the cross section will grow as 
\begin{equation}
\label{sigma}
\sigma \sim \sigma_{0} \, \frac{N^{N}}{N!}\,,
\end{equation}
namely is exponential increasing with the particles' multiplicity of the classicalon. 

Including now the factor determined in Eq.~\eqref{jaaj}, we obtain an extra exponential suppression in the cross-section, which matches the exponential factor at numerator, i.e.
\begin{equation}
\label{jajaj}
\sigma(2\rightarrow N_{Class})\sim M_{*}^{-2}e^{-N}\frac{N^{N}}{N!}\rightarrow M_{*}^{-2}\,. 
\end{equation}
The latter expression hence saturates the unitarity bound\footnote{A similar cross section behaviour was already observed in the case of pions, described by a similar non-linear sigma model \cite{pion1,pion2}} for $N\!>\!\!>\!1$ (in the well known Stirling approximation).

These considerations can be easily extended to other higher-order operators in our Lagrangian, as well as for the aforementioned $(\partial \phi)^{4}$ theory.

Another interesting process is the production of a classicalon through decays of massive particles. As well known, in standard quantum field theory a $1\rightarrow N$ decay process entails the amplitude
\begin{eqnarray}
\label{Gamma}
&&d\Gamma= \\
&&\frac{1}{M_{1}} \! \Big(\prod_{f}\frac{d^{3}p_{f}}{(2\pi)^{3}2E_{f}}\Big) \! |\mathcal{M}(m_{A}\rightarrow \{p_{f}\}|^{2}(2\pi)^{4}\delta^{4}\! (p_{A} \!-\! \sum p_{f} \! )\, , \nonumber
\end{eqnarray}
where $f$ denotes all N final states and $M_{1}$ the decaying particle mass. As explained above, the standard perturbative quantum field theory process is related to the $1\rightarrow classicalon$ transition, through an exponential phase --- see Eq.~\eqref{jaaj}. As noticed previously, the integration of the phase space scales as $E^{2N}/M_{*}^{2N}$, with the total energy equal to the decaying particle mass, 
i.e. $M_{1}^{2N}/M_{*}^{2N}$. Then, considering bosonic indinstinguibility, 
we must include a $N!$ factor on the denominator. Finally, the coherent state phase enters as $e^{-N}$, so that we infer the following decay rate: 
\begin{equation}
\label{Gamma}
\Gamma(1\rightarrow N_{Class})\sim \frac{M_{1}}{N!}e^{-N}\Big(c_{1}\frac{M_{1}}{M_{*}}\Big)^{2N}\sim \Gamma_{0}e^{-N+c_{1}\frac{M_{1}^{2}}{M_{*}^{2}}}\,. 
\end{equation}
The decay rate then starts to be exponentially suppressed, with $N$ higher than a critical number value $\bar{N}\sim (M_{1}/M_{*})^{2}$, as naively expected.

\section{Production of the superfluid state from a cooling cascade} \label{cas}

Let us consider a process involving $N'+2$ scalar particles $\Phi$ of the type
\begin{equation}
p_{1},p_{2}\ \ \longrightarrow \ \ p_{1'},p_{2'},...,p_{N'-1},p_{N'}\,.  \label{ndecay}
\end{equation}
For $E_{CM}>\!\!>\Lambda'$, this process is catalyzed by the formation of the classicalon.
In the early Universe, the center of mass energy $E_{CM}\simeq T$, where $T$ is the thermal bath temperature. The final states are injected in the early Universe with an average kinetic energy lower than the initial CM energy/temperature. This amounts to a cool down of  the temperature of the early Universe by scattering processes. In particular, any two initial particles with energy $T$ that undergo a process like in \eqref{ndecay}, induce a decay in $N'$ new particles of lower energies, which then re-scatter in cascade's processes that cool down the temperature of the dark sector. The entire transition is described by the thermal/quantum superposition of all possible cascade's processes
\begin{eqnarray}
&p_{1'},p_{2'}\rightarrow p_{1''},p_{2''},...,p_{N''-1},p_{N''} \nonumber \\
& \dots  \nonumber \\
&p_{N'-1},p_{N'}\rightarrow p_{1^{N'}},p_{2^{N'}},...,p_{N^{N'}},p_{N^{N'}} \nonumber \\
& \dots \nonumber \\
&+{\rm permutations}\,, \nonumber \\
\end{eqnarray}
where for permutations we consider all the possible $1', i'$ scatterings with $i'=1',...,N'$, and so forth in the cascade's process. The total transition rate is given by
\begin{eqnarray}
\langle\{p_{1},...,p_{M}\}\rangle\rightarrow \langle \{p_{1},...,p_{N}\} \rangle\,,
\end{eqnarray}
describing the entire cascade of the initial thermal bath into soft quanta. The thermal/quantum average over all the possible channels is expected to be the $2\rightarrow N$ scattering
$$1,2\rightarrow 1', 2',...,N'\,,$$
where 
$$s_{12}=T^{2}, \qquad s_{ij}=\frac{T^{2}}{N^{'2}},\qquad i,j\neq 1,2\,.$$

This means that the temperature of the dark sector rapidly cools down to $T'\sim T/\sqrt{N}$. 
Now, the number of emitted particles with the same final energy cannot be infinite because a process producing only IR soft particles cannot be possible. Indeed, the total number of emitted particles is bounded by their own mass scale $m_{\Phi}$, namely
$$\sqrt{N}\sim \frac{T}{m_{\Phi}}.$$
Because of the non-linearity of the Lagrangian in Eq.(\ref{L}), one must consider an infinite set of non-perturbative vertices in the cascade, turning the complete calculation into an impossible task. 

The direct proportionality of $N$ with the temperature implies that the thermally averaged cross section $2\rightarrow N_{Class}$ recasts as
\begin{equation}
\label{crros}
\langle\sigma\rangle_{T}\sim M_{*}^{-2}\frac{1}{\Gamma(1+c_{3}\frac{T^{2}}{m_{\phi}^{2}})}e^{-c_{1}\frac{T^{2}}{m_{\phi}^{2}}}\Big(c_{2}\frac{T^{2}}{m_{\phi}^{2}}\Big)^{\Big(c_{2}\frac{T^{2}}{m_{\phi}^{2}}\Big)}\,,
\end{equation}
where the Euler gamma-function is introduced replacing the factorial, and $c_{1,2,3}$ are $O(1)$ constants.


As clarified in the Sec.~\ref{classic}, the system classicalizes through scattering processes in which the transferred momenta are greater than the cut-off scale $M_*$. In the case under scrutiny here, specified in Sec.~\ref{mod}, $M_*=\Lambda\simeq 1 {\rm meV}$ corresponds to the cut-off scale of the effective theory introduced by Berezhiani and Khoury in Ref.~\cite{Berezhiani:2015bqa}. Formation of classicalons is then triggered at distances well larger than the scale individuated by $L_*=1/M_*=1/\Lambda$. Following this line of thought, we may envisage that the production of a state of classicalons can be directly attained also in other stages of the Universe's expansion, when different energy scales are taken into account. For instance, considering the inflationary scenario, reheating can trigger formation of classicalons at an energy scale proportional to the reheating temperature dictated by the specific model taken into account. Classicalons that are formed will then evaporate at the end of reheating, through the very same classicalization processes we described, restoring the standard picture at later cosmic times.

\section{Boltzmann equations} \label{bo}
\noindent
 In this section we take into account the cosmological evolution of the $\Phi$ number density and the formation at late times of the condensate state of non-relativistic $\Phi$-particles.

The evolution of the number density is described by the Boltzmann's equation,
\begin{equation}
\label{Booo}
\frac{dn_{\Phi}}{dt}+3Hn_{\Phi}=\sum_{i}\Gamma_{i}\,,
\end{equation}
with 
$$\Gamma_{i}=\int f_{\Phi}(p_{1})f_{\Phi}(p_{2})d\sigma_{i}\,,$$ 
in which we focus on $i=1,2\rightarrow N$ processes, and we take $f_{\Phi}$ to be Bose-Einstein functions with a negligible chemical potential $$f_{\Phi,i}(p)=(e^{\beta E_{\Phi,i}(p)}-1)^{-1}.$$

Inside the $\Gamma_{i}$, for $E>m_{\Phi}$, the cross section ``classicalize" to a constant, as in Eq.~(\ref{crros}). Consequently, the integration over $1,2$ momenta with the Bose-Einstein distributions just trivializes to a $O(1)$ constant multiplying Eq.~(\ref{crros}). 

The cascade cooling process depends on the initial $T$ parameter. Essentially, this process starts with the first production of $\Phi$-fields in the early Universe. 
For example, if the $\Phi$ particles were produced after the inflaton reheating, a natural temperature scale would turn out to be of the order of $T_{1}\simeq T_{re}\simeq 10^{9}\, {\rm GeV}\,.$
However, an overproduction problem could be encountered here. The energy density of the $\Phi$-particles cannot exceed the cosmological critical density bound. Therefore, it must hold that
\begin{equation}
\label{mphi}
m_{\Phi}n_{\Phi}^{0}<\rho_{c}\,, 
\end{equation} 
in which $n_{\Phi}^{0}$ stands for the number density during recombination, and $\rho_{c}=1.053 75(13)\times 10^{-5}\,h^{2}\, {\rm GeV/cm}^{3}$ (according to last Planck data \cite{PDG}) denotes the cosmological critical density. \\

We notice that during the reheating, the decay time of the inflaton into classicalons scales as 
\begin{equation}
\label{scale}
\tau=kM_{1}^{-1}e^{+\Big(c_{2}\frac{T}{M_1}\Big)-c_{1}\frac{M_{1}^{2}}{M_{*}^{2}}}\, ,
\end{equation}
where $k$ depends on the specific couplings of the inflaton with $\Phi$, i.e. it is model dependent. For $\bar{N}\simeq c_{1}\frac{M_{1}^{2}}{M_{*}^{2}}+\log k$, the characteristic time of decay is very short, and the reheating of the inflaton into classicalons becomes very efficient.



\begin{figure}[t]
\centerline{ \includegraphics [height=5.1cm,width=0.9\columnwidth]{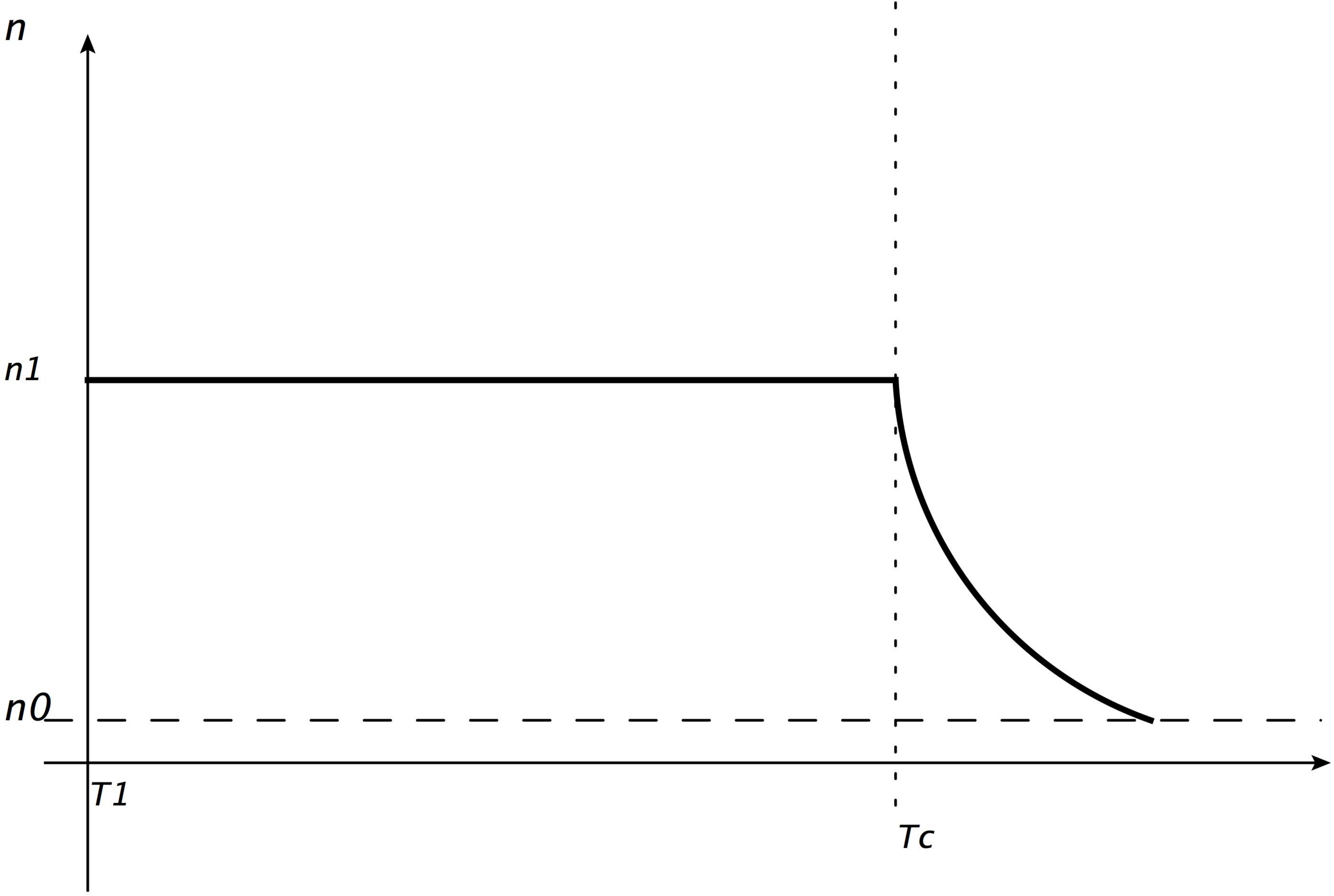}}
\vspace*{-1ex}
\caption{The evolution of the number density of $\Phi$ particles is displayed. From $T_{1}$ to $T_{c}$ the relativistic dilution factor is compensated by the cascade processes. On the other hand, after the phase transition at $T_{c}\simeq {1\div 10\,\rm eV}$ and the formation 
of the superfluid state, $\Phi$-particles are regarded as non-relativistic quanta of a classical state that is scaling as $T^{-3}$ along the Universe expansion. This means that after the critical temperature, the $\Phi$ field starts to behave as a cold dark superfluid.}
\label{plot}   
\end{figure}

The results of the $n$(T) evolution from the Boltzmann's equations are displayed in Fig.~3, having used the {\it ansatz} $\sigma=\langle \sigma(2\rightarrow N)\rangle_{\Phi_{\rm cl},T}$. We may give the following interpretation to the behavior that is shown. Le us consider two thermal epochs, namely $T_{1}>\!\!>m_{\Phi}$ and $T_{0}>\!\!>m_{\Phi}$, with $T_{0}<T_{1}$. During this thermal window, the $\Phi$ modes are strongly coupled each other and cannot freely propagate. Indeed, they are organized in classicalon states. This means that every $N\rightarrow N'$ process should be viewed as a thermally averaged classicalon-classicalon transition. Since the number of the $\Phi$ modes produced in each classicalon transition, namely $\langle N\rangle$, is proportional to the square of the average CM energy, in turn proportional to the square of the temperature, their number density scales with the Universe expansion as $n_{\Phi}\!\sim\! T^{5}$ in temperature. During this stage, particle are relativistic. We emphasize that this scaling is different from the one of free relativistic particles, $n_{\Phi}\!\sim\! T^{3}$. Therefore, the energy density of dark matter scales as $\rho=\rho_{0}\,T^{6}$.  \\

The right dark matter abundance can be obtained tuning the $k$-coupling of the inflaton to the classicalon. From the inflation energy scale, down to an energy scale few orders of magnitude above the particle mass, the dilution of the energy density scales as $\rho_{\Phi}/\rho_{I}\propto (T_{\Phi}/T_{I})^{6}$. Conservatively, we can estimate $T_{\Phi}$ as the temperature at which the inflaton decays in a coherent state that encodes a critical $N$ number of $\Phi$ particles, i.e. $T_{\Phi}=N\,m_{\Phi}$, still with $N\!>\!\!>\!1$. This means that at the reheating, where $\rho_{re}=k\rho_0$, the scaling can be found $\rho_{\Phi}/\rho_{I}=1/k \, [(N m_{\Phi})/T_{re}]^{6}$, where $T_{re}$ stands for the reheating temperature after inflation and $k$ is the inflaton coupling to the classicalon. If the number $N\!>\!\!>\!T_{re}/m_{\Phi}$, then the classicalon will be completely diluted and washed out, i.e. it cannot provide a good candidate of dark matter. There exists a critical number $N_{c}=T_{re}/m_{\Phi}\simeq 10^{18}$ at which the classicalon is no more diluted. This means that in the self-criticality phase, the production of quanta is compensated by the expansion of the Universe. Nonetheless, when the temperature cools down, new species of particles that compose the classicalon cannot be produced anymore. The classicalon will then cool down as cold superfluid dark matter, at smaller energy scales than the critical one at which $\Phi$ modes can be still be produced. In other words, the energy density will start scaling as $T^{3}$. \\

After this critical phase is reached, the DM energy-density can be estimated to be $\rho_{DM}\!\simeq\! k\rho_{\varphi}(T_{0}/T_{c})^{3}$, where $\rho_{\varphi}$ is the inflaton energy density, which we denote with $\rho_{\varphi}\simeq 3\,H_{I}^{2}/(8 \pi G)$, the temperature at the recombination epoch is $T_{0}$, and $T_{c}$ is the classicalon phase transition temperature. The energy-density of the inflaton field is enormous compared to other energy densities. Consequently, the dilution factor is very small, since the mass of $\Phi$ can be at least one order of magnitude larger the $T_{0}$. Roughly, the energy density of the inflaton can be estimated to be $\rho_{\phi}\simeq 10^{-1}M_{Pl}^{2}H_{I}\simeq 10^{36}\times 10^{20}\, {\rm GeV}^{4}$. On the other hand, the DM energy density is only $\rho_{\rm DM} \sim 0.1\rho_{c}\sim 10^{-6}{\rm GeV/cm}^{3}\sim 10^{-49}{\rm GeV}^{4}$, with $\rho_{c}$ critical energy density. This entails a large fine tuning of the coupling constant $k$, which is constrained to assume a value of about $10^{-105}$. A coupling so small can be understood in terms of effective operators of large dimension, such as $(1/\Lambda)^{N-3}\varphi\, \Phi^{N}$. Nonetheless, for $T_{c}\simeq 2m_{\Phi}\simeq 1\div 10\, {\rm eV}$, the $\Phi$ particles become non-relativistic, and the phase transition takes place. In other words, the $\Phi$ particles can softly form a superfluid state.

\section{Conclusions} \label{co}
\noindent
%
Superfluid dark matter was proposed in order to reconcile the intriguing success of MOND on galactic scales with the ${\rm \Lambda CDM}$ model, which is instead successful on cosmological scales. Within this scenario, dark matter consists of self-interacting axion-like condensates localized inside the galaxies. The superfluid phonons, weakly coupled to ordinary baryons, mediate an effective MOND acceleration. In superfluid dark matter, the effective MOND gravitational potential of a galaxy is different from the one of a galaxy cluster. Indeed, dark matter has a higher temperature in clusters, and a sizable part of axion-like particles are de-confined from the condensate state. Such a model was engineered in a phenomenologically healthy, non-relativistic effective, formulation. 

The problem of the UV completion of this model was not solved yet. In this paper, we have shown that a Wilsonian UV completion is not necessary for this model, and that superfluid dark matter ``self-unitarizes'' itself, undergoing ``classicalization" phenomena. Classicalization is an alternative to the Wilsonian UV completion, in which the formation of classical non-perturbative configuration, at a certain critical energy scale, unitarizes the scattering amplitudes of the classicalizing theory. In the case of superfluid dark matter, the classicalization mechanism can be directly related to the production of the superfluid condensate. In other words, the UV self-completion of the model is also related to another unsolved issue, namely how superfluid dark matter was generated in our Universe. In the model under scrutiny, the critical scale for the formation of the classical state is of about the {\rm meV}-scale, related to the cosmological constant scale. This means that every axion-like particle, which undergoes scatterings until the very late Universe, will form a classical state. The classical state will evaporate in a large number of soft axion-like particles. Nonetheless, this unitary process can efficiently happen at energies that are very far from the mass scale of the axion-like particle. In other words, this can happen if the average center of mass energy (the temperature of the Universe) is much larger than the axion-like mass. In order to fit correctly the galactic rotational curves, an axion-like mass of about $1-10\,{\rm eV}$ is necessary. This basically means that at energies larger than $1-10\,{\rm eV}$, but close to the recombination epoch, the classical state cannot evaporate into soft axion-like particles. Such a process exactly corresponds the formation of a superfluid dark matter condensate. This provides a new dark matter formation mechanism, related to classicalization, that is alternative to standard ones, as for instance WIMP thermal productions. \\

\acknowledgements 
\noindent
We acknowledge the anonymous Referee for useful comments that helped to clarify the content of this paper. 
We also would like to thank Alexander Vikman for interesting discussions on classicalization and superfluid dark matter.


\begin{thebibliography}{99}


\bibitem{Khoury:2014tka}
  J.~Khoury,
  Phys.\ Rev.\ D {\bf 91} (2015) no.2,  024022
  doi:10.1103/PhysRevD.91.024022
  [arXiv:1409.0012 [hep-th]].


\bibitem{Berezhiani:2015bqa}
  L.~Berezhiani and J.~Khoury,
  Phys.\ Rev.\ D {\bf 92} (2015) 103510
  doi:10.1103/PhysRevD.92.103510
  [arXiv:1507.01019 [astro-ph.CO]].


\bibitem{Berezhiani:2015pia}
  L.~Berezhiani and J.~Khoury,
  Phys.\ Lett.\ B {\bf 753} (2016) 639
  doi:10.1016/j.physletb.2015.12.054
  [arXiv:1506.07877 [astro-ph.CO]].


\bibitem{Dvali:2010jz}
  G.~Dvali, G.~F.~Giudice, C.~Gomez and A.~Kehagias,
  JHEP {\bf 1108} (2011) 108
  doi:10.1007/JHEP08(2011)108
  [arXiv:1010.1415 [hep-ph]].
  
\bibitem{Dvali:2012mx}
  G.~Dvali and C.~Gomez,
  JCAP {\bf 1207} (2012) 015
  doi:10.1088/1475-7516/2012/07/015
  [arXiv:1205.2540 [hep-ph]].
  
\bibitem{Bajc:2011ey}
  B.~Bajc, A.~Momen and G.~Senjanovic,
  arXiv:1102.3679 [hep-ph].
  
\bibitem{Percacci:2012mx}
  R.~Percacci and L.~Rachwal,
  Phys.\ Lett.\ B {\bf 711} (2012) 184
  doi:10.1016/j.physletb.2012.03.073
  [arXiv:1202.1101 [hep-th]].
  
\bibitem{Vikman:2012bx}
  A.~Vikman,
  EPL {\bf 101} (2013) no.3,  34001
  doi:10.1209/0295-5075/101/34001
  [arXiv:1208.3647 [hep-th]].
  
\bibitem{Addazi:2015ppa}
  A.~Addazi,
  Int.\ J.\ Mod.\ Phys.\ A {\bf 31} (2016) no.04n05,  1650009
  doi:10.1142/S0217751X16500093
  [arXiv:1505.07357 [hep-th]].
  
\bibitem{Alexander:2011hz} 
  S.~Alexander, A.~Marciano and D.~Spergel,
  JCAP {\bf 1304}, 046 (2013)
  doi:10.1088/1475-7516/2013/04/046
  [arXiv:1107.0318 [hep-th]].


\bibitem{Alexander:2014uza} 
  S.~Alexander, D.~Jyoti, A.~Kosowsky and A.~Marciano,
  JCAP {\bf 1505}, 005 (2015)
  doi:10.1088/1475-7516/2015/05/005
  [arXiv:1408.4118 [hep-th]].
  
\bibitem{Dona:2016fip} 
  P.~Dona and A.~Marciano,
  Phys.\ Rev.\ D {\bf 94}, no. 12, 123517 (2016)
  doi:10.1103/PhysRevD.94.123517
  [arXiv:1605.09337 [gr-qc]].
  
\bibitem{Addazi:2016rnz} 
  A.~Addazi, S.~Alexander, Y.~F.~Cai and A.~Marciano,
  arXiv:1612.00632 [gr-qc].
  
\bibitem{Brahma:2016wvk} 
  S.~Brahma, P.~Dona and A.~Marciano,
  arXiv:1612.00760 [gr-qc].
  
\bibitem{Alexander:2016xbm} 
  S.~Alexander, A.~Marciano and Z.~Yang,
  arXiv:1602.06557 [hep-th].
  
\bibitem{Addazi:2016nok} 
  A.~Addazi, A.~Marciano and S.~Alexander,
  arXiv:1603.01853 [gr-qc].
  
\bibitem{Alexander:2009yb} 
  S.~H.~S.~Alexander,
  arXiv:0911.5156 [hep-ph].
  
  \bibitem{Addazi:2016oob}
  A.~Addazi, S.~Capozziello and S.~Odintsov,
  Phys.\ Lett.\ B {\bf 760} (2016) 611
  doi:10.1016/j.physletb.2016.07.047
  [arXiv:1607.05706 [gr-qc]].

\bibitem{Dvali:2012en} 
  G.~Dvali and C.~Gomez,
  Eur.\ Phys.\ J.\ C {\bf 74}, 2752 (2014)
  doi:10.1140/epjc/s10052-014-2752-3
  [arXiv:1207.4059 [hep-th]].
  
  \bibitem{pion1}
F.~P.~Ermolov, E.~S.~Kokoulina, E.A.~Kuraev, A.Y.~Kutov, V.A.~Nikitin, A.A.~Pankov, I.A.~Roufanov, N.K.`Zhidkov,  
{\it Study of multiparticle production by gluon dominance model. Part II};
17th International Baldin Seminar on High Energy Physics Problems: Relativistic Nuclear Physics and Quantum Chromodynamics (ISHEPP 2004)

\bibitem{pion2}
  E.~S.~Kokoulina,
  arXiv:1504.01468 [hep-ph].
  
\bibitem{PDG}
pdg.lbl.gov/2012/reviews/rpp2012-rev-astrophysical-constants.pdf


\end{thebibliography}
\end{document}